\documentclass[twocolumn,secnumarabic,amssymb, nobibnotes,superscriptaddress, aps, prd]{revtex4-2}

\usepackage{graphicx}
\begin{document}

\title{Confined dipole and exchange spin waves in a bulk chiral magnet with Dzyaloshinskii-Moriya interaction}
\author{Ping Che}
\affiliation{Laboratory of Nanoscale Magnetic Materials and Magnonics, Institute of Materials (IMX), \'Ecole Polytechnique F\'ed\'erale de Lausanne (EPFL), 1015 Lausanne, Switzerland}
\author{Ioannis Stasinopoulos}
\affiliation{Physik Department E10, Technische Universit\"at M\"unchen, 85748 Garching, Germany}
\author{Andrea Mucchietto}
\affiliation{Laboratory of Nanoscale Magnetic Materials and Magnonics, Institute of Materials (IMX), \'Ecole Polytechnique F\'ed\'erale de Lausanne (EPFL), 1015 Lausanne, Switzerland}
\author{Jianing Li}
\affiliation{Laboratory of Nanoscale Magnetic Materials and Magnonics, Institute of Materials (IMX), \'Ecole Polytechnique F\'ed\'erale de Lausanne (EPFL), 1015 Lausanne, Switzerland}
\author{Helmuth Berger}
\affiliation{Institut de Physique de la Mati\`ere Complexe, \'Ecole polytechnique f\'ed\'erale de Lausanne (EPFL), 1015 Lausanne, Switzerland}
\author{Andreas Bauer}
\affiliation{Physik Department E51, Technische Universit\"at M\"unchen, 85748 Garching, Germany}
\author{Christian Pfleiderer}
\affiliation{Physik Department E51, Technische Universit\"at M\"unchen, 85748 Garching, Germany}
\author{Dirk Grundler}\email{dirk.grundler@epfl.ch}
\affiliation{Laboratory of Nanoscale Magnetic Materials and Magnonics, Institute of Materials (IMX), \'Ecole Polytechnique F\'ed\'erale de Lausanne (EPFL), 1015 Lausanne, Switzerland}
\affiliation{Institute of Microengineering (IMT), \'Ecole Polytechnique F\'ed\'erale de Lausanne (EPFL), 1015 Lausanne, Switzerland}
\date{\today}%

\begin{abstract}
The Dzyaloshinskii-Moriya interaction (DMI) has an impact on excited spin waves in the chiral magnet Cu$_2$OSeO$_3$ by means of introducing asymmetry on their dispersion relations. The confined eigenmodes of a chiral magnet are hence no longer the conventional standing spin waves. Here we report a combined experimental and micromagnetic modeling study by broadband microwave spectroscopy we observe confined spin waves up to eleventh order in bulk Cu$_2$OSeO$_3$ in the field-polarized state. In micromagnetic simulations we find similarly rich spectra. They indicate the simultaneous excitation of both dipole- and exchange-dominated spin waves with wavelengths down to (47.2 $\pm$ 0.05) nm attributed to the exchange interaction modulation. Our results suggest DMI to be effective to create exchange spin waves in a bulk sample without the challenging nanofabrication and thereby to explore their scattering with noncollinear spin textures.
\end{abstract}

\maketitle

\section{Introduction}

Magnon band structures are nontrivial in chiral magnets because of bulk Dzyaloshinskii-Moriya interaction (DMI) \cite{Kugler2015,Portnichenko2016_helixbandstructure,Roldan-Molina2016,Garst2017_review,Weber2018_SW_DMI_MnSi,Luo2020}. As a consequence of the asymmetric exchange interaction, bulk DMI introduces non-reciprocity for the spin waves \cite{Kataoka1987,Cortes2013,Seki2016_DMIinCSO, Sato2016_SW_DMI_MnSi, Weber2018_SW_DMI_MnSi,Weber2018,Weber2019_neuSca_MnSi,Seki2020_SWalongSKL}. Therefore chiral magnets can serve as non-reciprocal microwave devices \cite{Seki2016_DMIinCSO}. In order to make use of the nontrivial magnonic properties, finite wave-vector $k$ excitations in chiral magnets are strongly demanded. The spin dynamics in bulk chiral magnets with lateral dimension up to millimeters have been investigated experimentally \cite{Onose2012_FMR_CSO,Schwarze2015_NM,Okamura2013_RFmagnetoelectric, Okamura2015_Magnetochiral_Dichroism,Weiler2017_higherorderQ,Stasinopoulos2017_linear_Dichroism,Pollath2019_x-rayforFMR,Aqeel2020} but mostly focusing on a wave vector $k=0$. To explore the spin wave excitation with finite $k$, focused ion beam patterning was utilized to shape the chiral magnets into lamella and to flip them onto microstructured coplanar waveguides (CPWs) \cite{Seki2016_DMIinCSO,Seki2020_SWalongSKL}. Still, exchange-dominated spin waves in the gigahertz (GHz) frequency regime are largely unexplored.

In bulk magnets without DMI, spin waves exhibit symmetric dispersion relations for $+k$ and $-k$. Standing spin waves form with fixed nodes and antinodes. Intensities of corresponding spin wave resonances vary systematically with the order number (number of nodal lines) $n$ (supplementary Fig. S1) \cite{SSW_Grimsditch1979, SSW_Jorzick1999, SSW_Kampen2002_all_optical, SSW_Bayer2005,Demidov2007, SSW_Demidov2008,SSW_Mruczkiewicz2013,An2013,Bozhko2020}. In the chiral magnets with bulk DMI, the backward volume magnon dispersion relation is asymmetric [Fig.\ref{fig1_diagram} (a)] so that the spin waves propagating along opposite directions at the same frequency own different $k$. The discrepancy in $k$ does not allow for the conventional standing spin waves. Numerical methods showed involved spin waves phase profiles in nanoscale magnets with DMI \cite{Garcia2014,Zingsem2019,Jostem2020}. In thin films of the chiral magnet FeGe, an oscillating factor $\exp(-iQz)$ was introduced when the ferromagnetic resonance (FMR) at $k$ = 0 and perpendicular standing spin waves with finite $k$ were interpreted, where $Q$ is the pitch vector in a chiral magnet \cite{Zhang2017_FMR}. Standing spin waves were assumed in bulk $\rm{Cu_2OSeO_3}$ \cite{Weiler2017_higherorderQ,Aqeel2020} but their characteristics remained unexplored.

Here we report on confined spin waves observed in the bulk chiral magnet Cu$_2$OSeO$_3$ probed by broadband microwave spectroscopy and micromagnetic simulation. We excited spin waves by a dynamic field across the whole sample [Fig.\ref{fig1_diagram} (b)] and numerous resonance peaks with systematically varying intensities appeared in the field polarized phase [Fig.\ref{fig1_diagram} (c)]. We attributed those series of peaks to confined magnetostatic waves with wavevectors $\textbf{k}$ $\parallel$ $\textbf{H}$ and $\textbf{k}$ $\perp$ $\textbf{H}$. We performed micromagnetic simulations on a micron-sized sample exhibiting the same form factor, i.e., the same shape anisotropy (demagnetization effect). When considering a non-zero DMI, we observed both the odd and even order numbers $n$ of confined volume modes at low frequency consistent with the experimental observation. The simulation reveals the short-waved magnons down to a wavelength of about 40 nm coexisting at frequencies of the discretized volume modes. These modes depended characteristically on the DMI strength. The origin of the short-waved magnons in the exchange regime is attributed to the interplay of symmetric and asymmetric exchange interactions. They appear for both uniform and non-uniform excitation scenarios. Our results suggest DMI to be a strong tool to excite exchange-dominated spin waves without challenging nanofabrication.
\newpage

\section{Experimental technique and simulations}

\begin{figure}[h]
	\includegraphics[width=86mm]{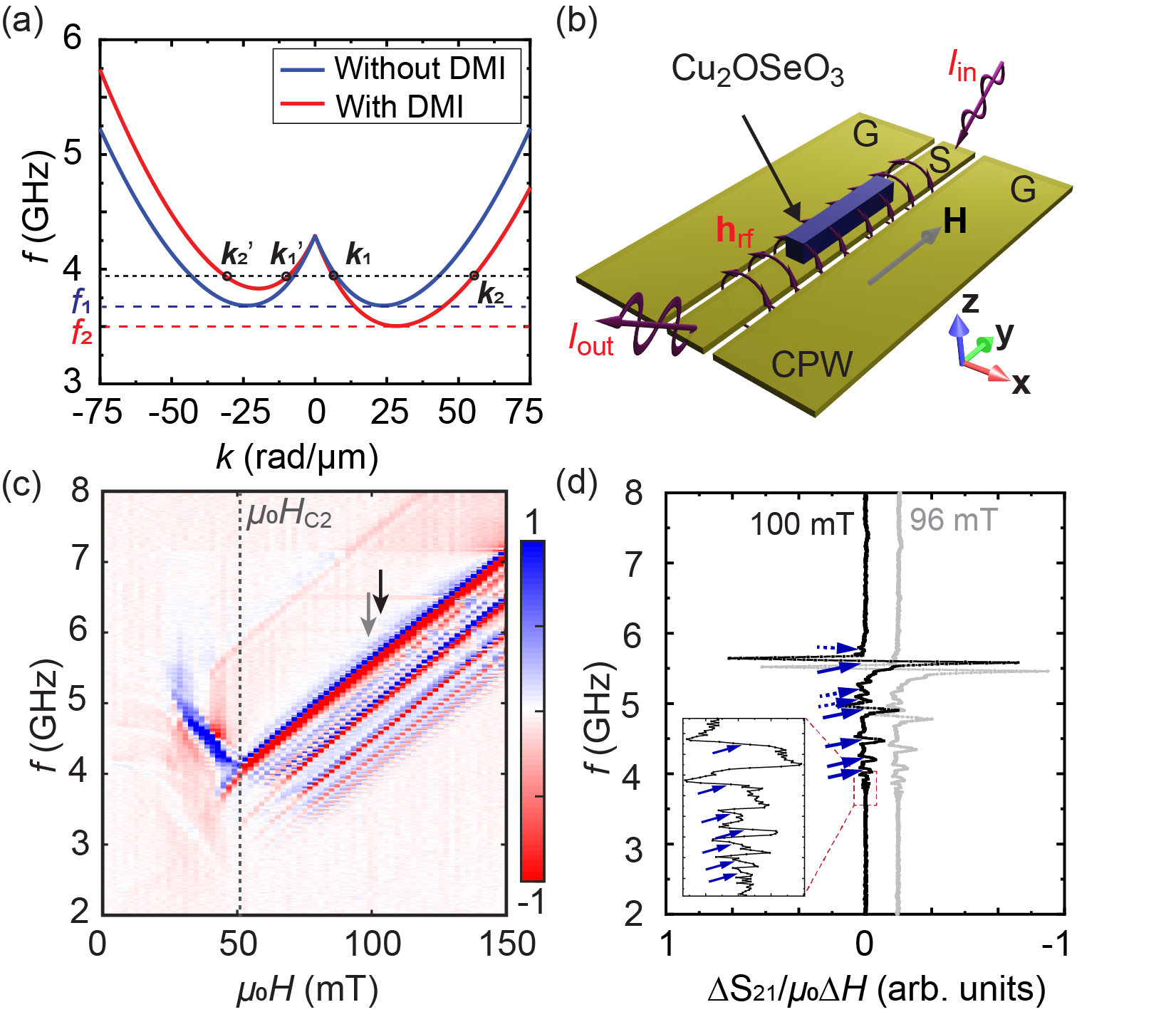}
	\caption{(a) Sketch of the dispersion relation of $\textbf{k}$ $\parallel$ $\textbf{H}$ mode with and without DMI. (b) Schematic diagram of bulk Cu$_2$OSeO$_3$ sample placed on CPW (yellow colored). (c) Color-coded maps of broadband spin wave spectra taken as a function of applied field $\mu_0 H$ along $\textbf{y}$ at 5 K. Color scale bar represents $\Delta S_{\rm 12}/\mu_0 \Delta H (\times 10^{-3}\;$T$^{-1})$. Black and gray arrows indicate where the lineplots in (d) were taken. (d) Lineplots of spectra $\Delta S_{\rm 12}/\mu_0 \Delta H$ at 5 K and $\mu_0 H$ = 100 mT (black) and 96 mT (gray) indicated by arrows in (c). Solid blue (and light blue) arrows mark the frequencies which we attribute to discrete $\textbf{k}$ $\parallel$ $\textbf{H}$ modes and dashed blue (and light blue) arrows mark the frequencies that we categorize in terms of discrete $\textbf{k}$ $\perp$ $\textbf{H}$ modes.}
	\label{fig1_diagram}
\end{figure}

The broadband microwave measurements were conducted on a bar-shaped Cu$_2$OSeO$_3$ sample. Cu$_2$OSeO$_3$ was reported to exhibit particularly low damping at low temperature \cite{Stasinopoulos2017_lowdamping}. The volume of the sample was 1 mm $\times$ 0.29 mm $\times$ 0.29 mm and all three surfaces were perpendicular to the easy axes ${\rm{[100]}}$, ${\rm{[010]}}$ and ${\rm{[001]}}$. It was placed on a commercial coplanar waveguide (CPW), which contained a 1 mm wide signal line and two ground lines as sketched in Fig. \ref{fig1_diagram} (b). A radio-frequency current $I_{\rm{rf}}$ was injected into the CPW by a vector network analyzer (VNA) and induced a dynamic magnetic field $\textbf{h}_{\rm{rf}}$ (purple circular arrows). The Cu$_2$OSeO$_3$ sample was positioned at the center of the CPW in a way that the longer axis of the sample was parallel to the signal line. Because the width of the signal line was 1 mm and more than three times wider compared to the sample width $\Delta x$ of 0.29 mm, the in-plane component of $\textbf{h}_{\rm{rf}}$ was assumed to be uniform with respect to $\textbf{x}$ and $\textbf{y}$ (it varied as a function of $\textbf{z}$) \cite{Stasinopoulos2017_linear_Dichroism}. Measurements were conducted at two cryogenic temperatures $T$: 5 K and 20 K (supplementary materials). Magnetic fields were applied by a superconducting magnet along $\textbf{y}$-direction. A residual magnetic field of 36 mT along $\textbf{z}$ was present due to remanence of the setup when the sample was cooled down. The reported field values $\mu_0 H$ describe the additional field applied via a current in the superconducting coil. The spectra discussed in this paper were taken at $\mu_0 H$ $\textgreater$ 96 mT, i.e., the applied field dominated the dynamic response. At temperature $T$ = 5 K, $\mu_0 H$ = 150 mT was first applied along $+\textbf{y}$ to saturate the Cu$_2$OSeO$_3$ and then reduced in a step-wise manner. The same saturation process was conducted when we changed the applied field direction to $\textbf{H}$ $\parallel$ $\textbf{x}$ and $\textbf{H}$ $\parallel$ $\textbf{z}$ for different measurements. The sample was heated to 20 K to perform the same series of measurements with the different field directions at different $T$.

In order to explore the role of DMI for confined spin waves, we conducted spin dynamics simulations using the micromagnetic software Mumax3 \cite{Mumax3} assuming $T=0$. A bar-shaped sample with 0.128 $\times$ 2.048 $\times$ 0.128 $\mu$m$^3$ was considered. The cell sizes were 4 nm along $\textbf{y}$ and 8 nm along $\textbf{x}$ and $\textbf{z}$ to resolve quantized spin waves. Here we used 1 $\leq$ ix $\leq$ 16, 1 $\leq$ iy $\leq$ 512 and 1 $\leq$ iz $\leq$ 16 to identify the cell number along $\textbf{x}$, $\textbf{y}$ and $\textbf{z}$ directions. Parameters such as exchange stiffness $A_s=7\times$ 10$^{-13}$ J/m, DMI constant $D$ = 7.4 $\times$ 10$^{-5}$ J/m, saturation magnetization $M_{\rm sat}$ = 1.03 $\times$ 10$^5$ A/m, 1st order cubic anisotropy constant $K_{\rm C1}$ = 6 $\times$ 10$^2$ J/m$^{3}$ and Gilbert damping constant $\alpha= 5\times 10^{-4}$ modelled Cu$_2$OSeO$_3$ and were taken from Refs. \cite{Stasinopoulos2017_lowdamping,Janson2014,Zhang2018}. In additional simulations, we varied the parameter $D$ between 0 and 10 ($\times$ 10$^{-5}$ J/m$^{2}$). The static magnetic field $\mu_0 H$ was varied from 160 mT to 80 mT along $\textbf{y}$ with 0.5 degree tilting angle with respect to $\textbf{x}$ to avoid numerical errors. The micromagnetic simulations do not consider the residual field in the experiment. The dynamic magnetic field $\mu_0 h$ in the format of a sinc function for broadband excitation was applied along $\textbf{x}$ and 0.5 degree tilted to $\textbf{z}$.

\section{Dispersion relation modified by bulk Dzyaloshinskii-Moriya interaction}

Before we present results, it is instructive to look into the dispersion relations modified by bulk DMI compared with the conventional case in Fig.\ref{fig1_diagram} (a) \cite{Cortes2013,KA1986}. In the ferrimagnetic state, the effective field term generated by bulk DMI enters the dispersion relation of only the spin wave mode with $\textbf{k}$ $\parallel$ $\textbf{H}$ and has no influence on the spin wave mode with $\textbf{k}$ $\perp$ $\textbf{H}$. With bulk DMI $D$ $\textgreater$ 0, the $\textbf{k}$ $\parallel$ $\textbf{H}$ mode dispersion relation is asymmetric with respect to $k$ = 0 rad/$\mu$m: the branch with positive $\boldmath k$ owns lower frequency than the system with no DMI and the branch with negative $\boldmath k$ owns higher frequency. In Fig. \ref{fig1_diagram} (a) the asymmetry due to DMI is exaggerated to make the difference in dispersion relations visible. When the bulk DMI is absent, the wave vectors at a given frequency own the same modulus of different sign so a standing spin wave with vector $|k_0|$ = ($|k_+|$ + $|k_-|$)/2 can form with fixed nodes and antinodes in a sample with finite length (width) $L$. In a one-dimensional case, $|k_0|$ is specified as $|k_0|$ = $n\pi$/$L$ where $n$ counts the number of nodes $0, 1, ..$ \cite{SSW_Grimsditch1979}. If no specific surface asymmetry plays a role, broadband spin wave spectroscopy based on a wide CPW can barely detect the odd modes because the antinodes with $\pi$ phase shift induce counter-acting voltage signals in the signal line of the CPWs which cancel each other. When bulk DMI is present, the dispersion relation is no longer symmetric so that at each frequency $|k_1| \neq |k_1^,|$ and $|k_2| \neq |k_2^,|$. Traveling waves with fixed nodes are expected to form but spin waves amplitudes have "a
pronounced time-dependent asymmetry in the mode profile" \cite{Zingsem2019}. In this scenario, the voltage signals induced by the odd modes do not cancel and one expects an induced net voltage in the CPW. When analyzing our experimental data, we discuss $|k_0|$ = ($|k_1|$ + $|k_1^,|$)/2 and not the individual values $|k_1|$ and $|k_1^,|$. This is because the excitation takes place at a fixed frequency and multiple $k$ ($|k_1|$ and $|k_1^,|$) are excited at the same time. In the micromagnetic simulations (methods), we analyze the integrated Fourier transform of spin wave amplitudes consistent with the signal detected by a CPW. With the help of the simulated phase profiles, we extract $|k_0|$ by a fitting procedure.


\section{Broadband spin wave spectroscopy data}

\begin{figure*}[htb]
	\centering
	\includegraphics[width=170mm]{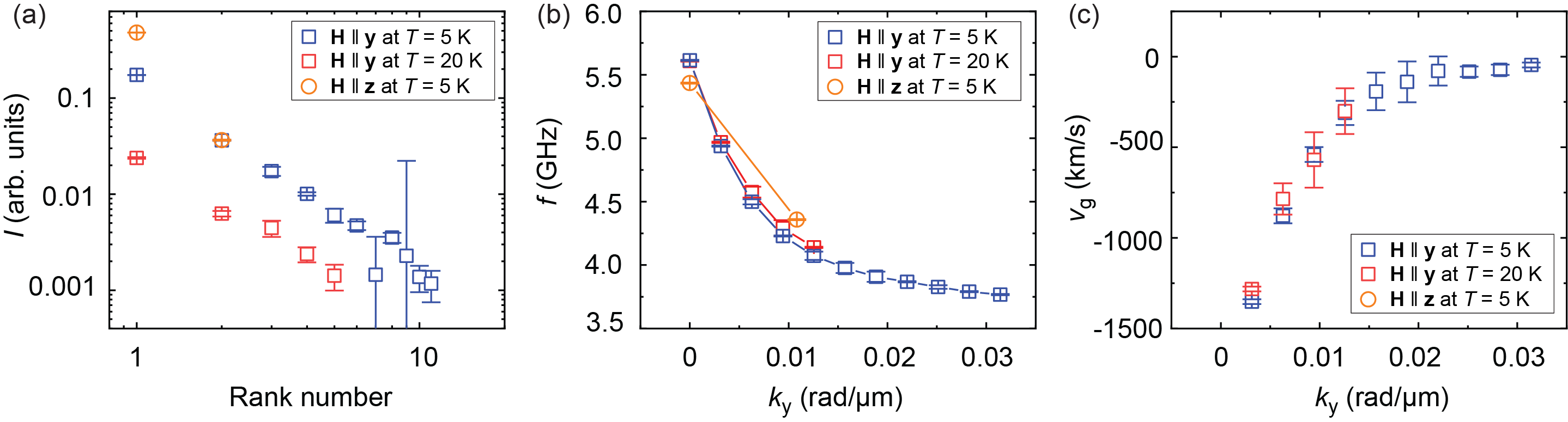}
	\caption{(a) Peak-to-peak intensity of resonances in a double-logarithmic plot corresponding to confined $\textbf{k}$ $\parallel$ $\textbf{H}$ mode sequences. Error bars represent the noise level in the spectra and the deviation in reading out the peaks. (b) Dispersion relation of spin waves extracted experimentally. Error bars reflect the frequency resolution of the VNA and the deviation in reading out the peaks. (c) Group velocities calculated as $2\pi \Delta f/\Delta k_y$ from (b). Error bars originate from the frequency uncertainty in (b). In (a), (b) and (c), blue squares were extracted from solid blue arrows in Fig. \ref{fig1_diagram} (d) of $\textbf{H}$ $\parallel$ $\textbf{y}$; red squares were extracted from data taken at $\mu_0 H$ = 112 mT at $T$ = 20 K (supplementary Fig. S2); yellow circles were extracted from data taken at $\mu_0 H$ = 150 mT at $T$ = 5 K when $\textbf{H}$ $\parallel$ $\textbf{z}$ (supplementary Fig. S3). Spin-wave dispersion relations evaluated from three different experiments are consistent in (b).}
	\label{fig2_dispersion}
\end{figure*}

Field dependent spectra taken at $T$ = 5 K are shown in Fig.~\ref{fig1_diagram} (c) for $\textbf{H}$ $\parallel$ $\textbf{x}$. A change in slope $df$/$dH$ of resonance frequencies $f$ was seen at $\mu_0 H_{\rm C2}$ = 52 mT marked by a black dashed line. For $\mu_0 H$ $\textgreater$ $\mu_0 H_{\rm C2}$, resonance frequencies $f$ increased almost linearly with $H$ indicating the field polarized phase. For $\mu_0 H$ $\textless$ $\mu_0 H_{\rm C2}$, the slope $df$/$dH$ indicated an unsaturated state, such as the conical phase in Cu$_2$OSeO$_3$ \cite{Schwarze2015_NM}. In the following we focus on the numerous resonances detected for $\mu_0 H$ $\textgreater$ $\mu_0 H_{\rm C2}$. Line spectra at fields $\mu_0 H$ = 100 mT and $\mu_0 H$ = 96 mT are shown in Fig.\ref{fig1_diagram} (d). 15 peaks are observed and marked by arrows. A color-coded map of spin waves spectra at $T$ = 20 K and corresponding line spectra at $\mu_0 H$ = 112 mT and $\mu_0 H$ = 108 mT are shown in the supplementary materials Fig. S2. At 20 K, 5 peaks (marked by arrows) were clearly visible above $\mu_0 H_{\rm C2}$. The small number of resonance is attributed to an increased damping, consistent with earlier reports on temperature-dependent damping \cite{Stasinopoulos2017_lowdamping,Seki2020_SWalongSKL}.

We have categorized the observed resonances depending on their systematic peak-to-peak intensity variation. Solid arrows mark $\textbf{k}$ $\parallel$ $\textbf{H}$ modes with different wave vectors $k_y$. The sequence of peaks was ranked with order numbers 1, 2, ..., with intensity $I$ varying from high to low. The intensities $I$ of resonances marked by solid arrows are summarized in Fig.\ref{fig2_dispersion} (a) in a double-logarithmic manner. The linear variation suggests that $I$ is inversely proportional to the order number which is a sign of confined spin waves with different numbers of nodes \cite{An2013}. Considering $\textbf{k}$ $\parallel$ $\textbf{H}$ modes to be confined along $\textbf{y}$ we attribute wave vectors to the discrete modes of Fig. \ref{fig1_diagram} (d) according to $k_y$ = ($\pi n_{\rm y}$)/${\Delta y}$ where $n_{\rm y} = 0, 1, ...$ and ${\Delta y} =L= 1$~mm is the length of the bar-shaped sample along the magnetic field direction. Figure \ref{fig2_dispersion} (b) now shows the resonance frequencies identified in Fig. \ref{fig1_diagram} (d) as a function of estimated values $k_y$ (blue squares). The frequencies follow $f(k)$ expected for $\textbf{k}$ $\parallel$ $\textbf{H}$ modes. The group velocities $v_{\rm g}$ calculated from (b) according to $v_{\rm g}=2\pi\Delta f /\Delta k_y$ are shown in Fig. \ref{fig2_dispersion} (c). We find a value $v_{\rm g}$ of -1300 km/s near $k=0$. This value reflects a backward volume magnetostatic wave in the long-wavelength limit in the dipolar regime of a relatively thick ferrimagnet.

Dashed arrows in Fig.\ref{fig1_diagram} (d) mark additional discretized $\textbf{k}$ $\perp$ $\textbf{H}$ modes formed on top of $\textbf{k}$ $\parallel$ $\textbf{H}$ modes. The combined wave vector of such excitations reads $k^2={k_y}^2+{k_\perp}^2$ where $k_\perp$ represents the wave vector in a direction transverse to the magnetic field vector \cite{Stancil_book}. The peaks marked by dash arrows follow the sequence of conventional standing spin waves because the bulk DMI does not modify the dispersion relation of the spin waves when $\textbf{k}$ $\perp$ $\textbf{H}$ and those modes exhibit a positive group velocity. Hence higher-order confined modes exhibit higher resonance frequencies. $\textbf{k}$ $\perp$ $\textbf{H}$ modes confined along $\textbf{z}$ and $\textbf{x}$ are degenerate because the dimensions of the considered bulk Cu$_2$OSeO$_3$ sample amount to $\Delta z$ = $\Delta x$ = 0.29 mm. 

The colored-coded maps of broadband spin wave spectra for other field directions are plotted in the supplementary Fig. S3. When the field was applied along $\textbf{z}$, only one further quantized $\textbf{k}$ $\parallel$ $\textbf{H}$ mode was resolved at lower frequency because the wave vector determined by the dimension $\Delta z$ = 0.29 mm is much larger and may approach the bottom of the $\textbf{k}$ $\parallel$ $\textbf{H}$ magnetostatic wave band. The dispersion relations and group velocities are consistent for the configuration of $\textbf{H}$ $\parallel$ $\textbf{z}$ [Fig. \ref{fig2_dispersion} (b) and (c)]. When the field was applied along $\textbf{x}$, a confined $\textbf{k}$ $\parallel$ $\textbf{H}$ mode was not resolved, most likely due to a vanishing torque because of $\textbf{h}_{\rm{rf}}$ $\parallel$ $\textbf{M}$. Still the FMR and edge modes were observed. The intensities are low.

\section{Simulated modes and discussion}

\begin{figure}
	\includegraphics[width=86mm]{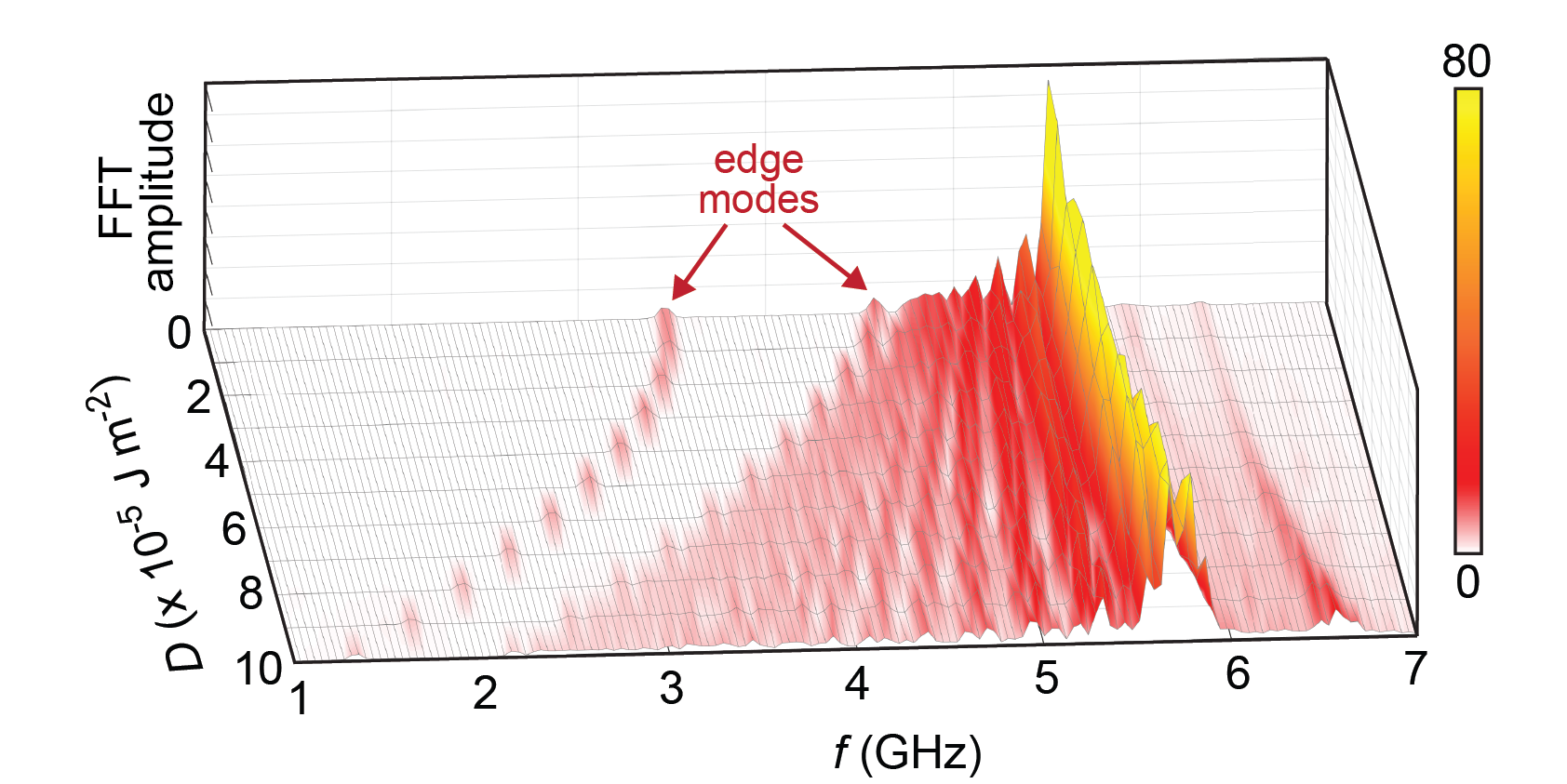}
	\caption{Spin-waves spectra for different DMI constants. The color bar represents the integrated Fourier transform (FFT) amplitude of magnetization component $m_{\rm z}$.}
	\label{fig3_simulation_DMI_noDMI}
\end{figure}

\begin{figure}
	\includegraphics[width=86mm]{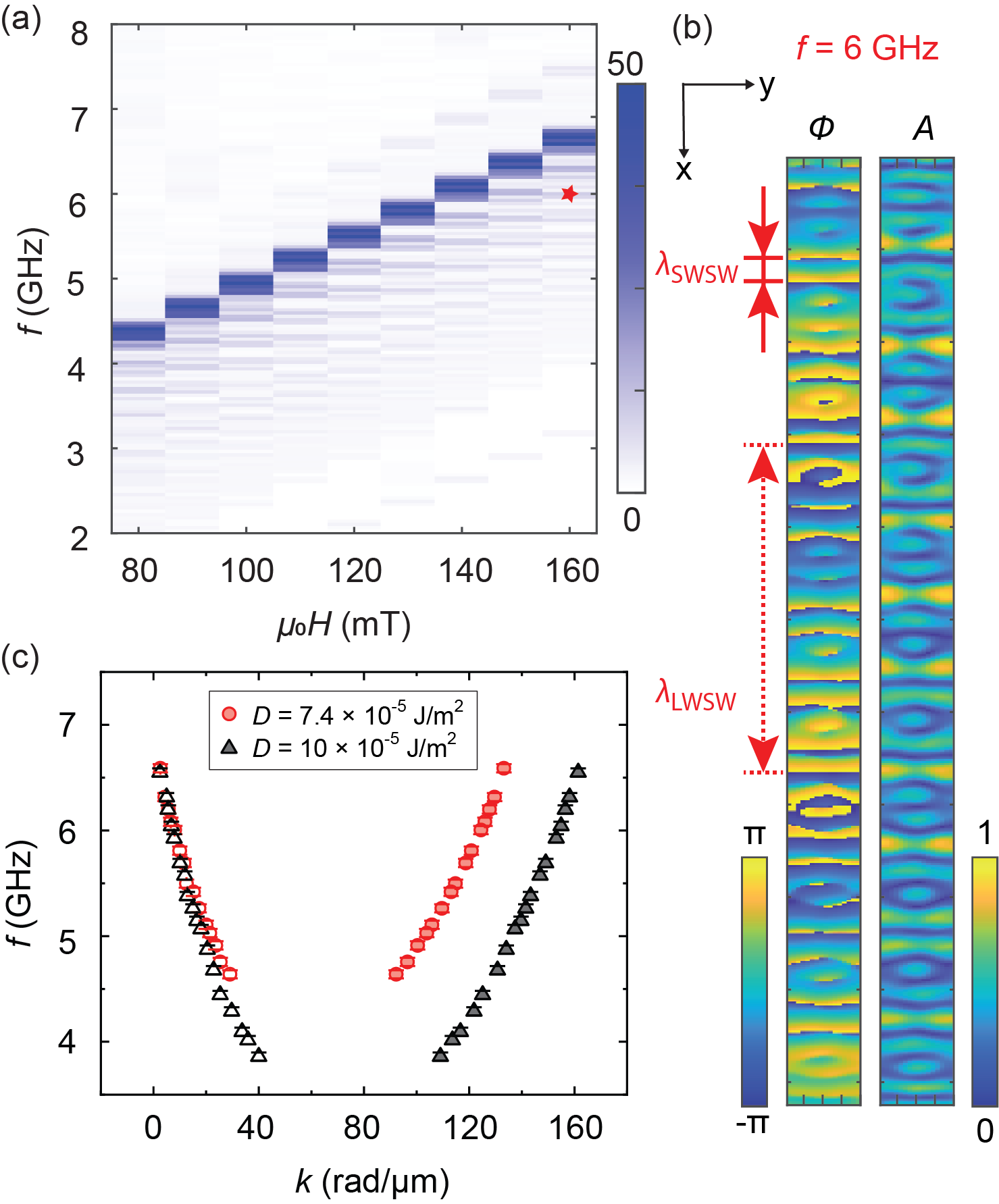}
	\caption{(a) Color-coded maps of spin wave spectra from micromagnetic simulations in layer 6 in $\textbf{z}$ direction. Color scale bar represents the amplitude of Fourier transform of $m_z$. (b) Color coded maps of amplitude $A$ and phase $\it{\Phi}$ in $x-y$ plane of $z$ layer iz = 6, with $D$ = 7.4 $\times$ 10$^{-5}$ J/m$^{2}$ at the field value indicated in (a) by the red star. The color scale of amplitude maps are normalized so that blue represents 0 and yellow represents 1. In the phase maps, blue and yellow color represents -$\pi$ and $\pi$, respectively. The wavelength of the SWSW and LWSW were marked. (c) Dispersion relation of resonance frequencies extracted from the micromagnetic simulations. Red squares are taken from $\mu_0 H$ = 160 mT with $D$ = 7.4 $\times$ 10$^{-5}$ J/m$^{2}$. Black triangles are taken from $\mu_0 H$ = 160 mT with $D$ = 10 $\times$ 10$^{-5}$ J/m$^{2}$. Open symbols are LWSWs and correspond to the dashed red arrow in (b). Filled symbols are SWSWs and correspond to the solid red arrows in (b). }
	\label{fig4_simulation_dispersion_relation}
\end{figure}

The integrated Fourier transform (FFT) amplitude of magnetization components $|m_{\rm z}|$ of all cells is plotted in Fig. \ref{fig3_simulation_DMI_noDMI} and Fig. \ref{fig4_simulation_dispersion_relation} (a). In both data sets we find one mode of largest intensity. On the low-frequency side of this most prominent peak, multiple resonances peaks are seen. This is consistent with the observation in the experiment [Fig. \ref{fig1_diagram} (c)]. It is noted that when the PBC was applied along $\textbf{y}$, there were no multiple peaks at lower frequency in the simulation. Therefore, the resonance peaks at lower frequency reflected discrete spin waves confined by the sample boundaries in $y$-direction.

The bulk DMI constant was varied from 0 to 10 ($\times$ 10$^{-5}$ J/m$^{2}$) in Fig. \ref{fig3_simulation_DMI_noDMI}. Here the applied field was fixed at 160 mT. Multiple peaks with decreasing amplitudes towards lower frequency are present for all the spectra, substantiating the formation of discrete modes due to spin wave confinement between sample boundaries for all the different values $D$. When $D = 0$, confined spin waves reside between 4.4 GHz and 7.3 GHz and the highest intensity peak appears at 6.5 GHz. The discrete $\textbf{k}$ $\parallel$ $\textbf{H}$ mode resonances reflect numbers $n_y=1, 3, 5,...$. On the one hand, when $D$ increases up to 10 $\times$ 10$^{-5}$ J/m$^{2}$, the peak with the highest intensity moves to a higher frequency of 6.8 GHz and more $\textbf{k}$ $\parallel$ $\textbf{H}$ mode resonance peaks appear down to 2.3 GHz (instead of 4.4 GHz). As shown in Fig. \ref{fig1_diagram} (a), when bulk DMI is present, the dispersion relation $f(k)$ of $\textbf{k}$ $\parallel$ $\textbf{H}$ magnetostatic waves is asymmetric. This asymmetry increases with $D$. For $D$ $\neq$ 0, the lowest frequency $f_2$ possible to be excited is lower than $f_1$ which denotes the bottom of the symmetric $\textbf{k}$ $\parallel$ $\textbf{H}$ mode band in case of $D$ = 0. The lower frequency $f_2$ explains the larger frequency regime in which discrete $\textbf{k}$ $\parallel$ $\textbf{H}$ modes occur for increasing $D$. On the other hand, the asymmetry results in the occurrence of, both, the even and odd-numbered modes $n_y$ in the integrated FFT amplitudes of our simulation. Thus all the confined $\textbf{k}$ $\parallel$ $\textbf{H}$ modes with $n_y=1, 2, 3, ...$ appear and the total number of resonance peaks increases with $D$.

For $f$ = 6 GHz and $D$ = 7.4 $\times$ 10$^{-5}$ J/m$^{2}$ we plot the simulated spin-precessional amplitudes $A$ and phases $\it{\Phi}$ inside the chiral magnet (for layer iz = 6) in Fig. \ref{fig4_simulation_dispersion_relation}(b). In both the $A$ and $\it{\Phi}$ maps, a confined long-waved (marked by a dashed red arrow) and short-waved spin wave (marked by solid red arrows) are resolved. The phase line plots for iz = 6 and ix = 6 of all the resonances at 160 mT are plotted in the supplementary Fig. S4. All of them display the coexistence of two waves with different wavelengths. We label them as long-waved spin waves (LWSWs) and short-waved spin waves (SWSWs) in the following.


The simulated resonance frequencies of both LWSWs and SWSWs are summarized in Fig. \ref{fig4_simulation_dispersion_relation} (c). The open symbols representing the LWSWs follow the dispersion relation of the dipolar $\textbf{k}$ $\parallel$ $\textbf{H}$ magnetostatic waves with a negative group velocity consistent with the experimental results shown in Fig. \ref{fig2_dispersion}(b). It is notable that the wavevectors of LWSWs do not match perfectly with the calculation in experimentally extracted data as $k_y$ = ($\pi n_{\rm y}$)/${\Delta y}$ ($n_{\rm y} = 0, 1, ...$ and  ${\Delta y}$ is the sample length). We interpret the reason to be the inhomogeneity of the internal field in the simulated sample near the edges which modifies the wave format. It is not counted in the calculation of wavevectors in Fig. \ref{fig2_dispersion} because the sample dimension along $\textbf{y}$ in experiment is as large as $y$ = 1 mm and thus the influence of inhomogeneous region is negligible. The filled symbols representing SWSWs show the characteristics of exchange-dominated spin waves with a positive group velocity. In case of $D$ = 7.4 $\times$ 10$^{-5}$ J/m$^{2}$ modelling Cu$_2$OSeO$_3$ the largest resolved wave vector amounts to $k$ = 133 rad/$\mu$m corresponding to a wavelength $\lambda$ = (47.2 $\pm$ 0.05) nm. When we increase the DMI constant to $D$ = 10 $\times$ 10$^{-5}$ J/m$^{2}$, the maximum wave vector of the exchange-dominated SWSWs increases to $k$ = 162 rad/$\mu$m, corresponding to $\lambda$ = (38.9 $\pm$ 0.04) nm. However, the wavelengths of the LWSWs remain similar to the smaller DMI value. The small deviation is because of the enhancement of the asymmetry of the $\textbf{k}$ $\parallel$ $\textbf{H}$ mode dispersion relation. The sign of DMI does not play a role for the extracted absolute wave vector values but it changes the sign of phase velocity of both LWSWs and SWSWs [supplementary Fig. S6]. We attribute the coexistence of multiple wave vector excitations at each frequency in the simulations to the exchange interactions. The asymmetric DMI, together with the symmetric exchange interaction, creates a periodic modulation of exchange interactions for spins in chiral magnets and the period is controlled by the strength of DMI constant $D$ and exchange stiffness $A$. When the confined $\textbf{k}$ $\parallel$ $\textbf{H}$ modes with $k_1$ are excited because of the sample boundaries, the modulation of exchange interactions provide the source of $k_2$ excitation at the same frequency.


\section{Conclusion}

In summary, we reported confined magnetostatic waves formed by the sample boundaries in a bar-shaped bulk $\rm{Cu_2OSeO_3}$ sample explored by both broadband spin wave spectroscopy and micromagnetic simulations. In the simulations ultrashort spin waves down to (47.2 $\pm$ 0.05) nm were predicted beyond the experimentally observed dipolar spin wave modes with $\textbf{k}$ $\parallel$ $\textbf{H}$. They were attributed to the DMI induced asymmetry of the dispersion relation and the periodic modulation of exchange interactions in chiral magnets. It has been proved that the wavelength and phase velocity strongly depended on the DMI strength. By increasing the DMI strength, shorter wavelengths have been achieved. Our findings provide an alternative way of exchange-dominated spin wave excitation without the need of nanofabrication. 

\section*{Acknowledgement}

The authors thank Prof. Markus Garst from Karlsruhe Institute of Technology for the discussions and comments on the manuscript. We acknowledge the financial supports from Swiss National Science Foundation (SNSF) Sinergia Network NanoSkyrmionics CRSII5 171003, Deutsche Forschungsgemeinschaft (DFG, German Research Foundation) under TRR80 (From Electronic Correlations to Functionality, Project No. 107745057, Projects E1 and F7), SPP2137 (Skyrmionics, Project No. 403191981, Grant PF393/19), and the excellence cluster MCQST under Germany's Excellence Strategy EXC-2111 (Project No. 390814868). Financial support by the European Research Council (ERC) through Advanced Grants No. 291079 (TOPFIT) and No. 788031 (ExQuiSid) is gratefully acknowledged.

\end{document}